\documentclass[twocolumn,superscriptaddress,aps,preprintnumbers,amsmath,amssymb,prl,nofootinbib]{revtex4}

\usepackage{amsmath}
\usepackage{amssymb}
\usepackage{amsfonts}
\usepackage{graphicx}
\usepackage{xcolor}
\usepackage{xfrac}
\usepackage{comment}
\usepackage{pifont}
\usepackage{physics}
\usepackage{fourier}
\usepackage{hyperref}
\usepackage{bm}
\usepackage{enumitem}

\definecolor{rossoferrari}{HTML}{D9073D}
\definecolor{mediumblue}{HTML}{0000CD}
\definecolor{forestgreen}{HTML}{228B22}
\definecolor{desy_blue}{HTML}{009EE2}
\definecolor{desy_orange}{HTML}{FD8800}
\definecolor{light_pink}{rgb}{1,0.4,0.4}
\definecolor{light_blue}{rgb}{0.284602,0.317763,0.963947}
\hypersetup{setpagesize=false,bookmarksnumbered=true,bookmarksopen=true,%
colorlinks=true,linkcolor=light_blue,urlcolor=rossoferrari,citecolor=rossoferrari,linktocpage=false}

\newcommand{\MeV}{\,\mathrm{MeV}}
\newcommand{\GeV}{\,\mathrm{GeV}}

\begin{document}


\preprint{CERN-TH-2022-134}
\preprint{RESCEU-13/22}
\preprint{KEK-TH-2441}
\preprint{MS-TP-22-23}
\preprint{TU-1164}

\title{A new constraint on primordial lepton flavour asymmetries}

\author{Valerie Domcke}
\email{valerie.domcke@cern.ch}
\affiliation{Theoretical Physics Department, CERN, 1211 Geneva 23, Switzerland}
\affiliation{Institute of Physics, EPFL, 1015 Lausanne, Switzerland}

\author{Kohei Kamada}
\email{kohei.kamada@resceu.s.u-tokyo.ac.jp}
\affiliation{Research Center for the Early Universe, The University of Tokyo, Hongo 7-3-1 Bunkyo-ku, Tokyo 113-0033, Japan}

\author{Kyohei Mukaida}
\email{kyohei.mukaida@kek.jp}
\affiliation{KEK Theory Center, Tsukuba 305-0801, Japan}
\affiliation{Graduate University for Advanced Studies (Sokendai), Tsukuba 305-0801, Japan}

\author{Kai Schmitz}
\email{kai.schmitz@uni-muenster.de}
\affiliation{University of M\"unster, Institute for Theoretical Physics, 48149 M\"unster, Germany}

\author{Masaki Yamada}
\email{m.yamada@tohoku.ac.jp}
\affiliation{FRIS, Tohoku University, Sendai, Miyagi 980-8578, Japan}
\affiliation{Department of Physics, Tohoku University, Sendai, Miyagi 980-8578, Japan}

\date{\today}


\begin{abstract}
\noindent
A chiral chemical potential present in the early universe can source helical hypermagnetic fields through the chiral plasma instability. If these hypermagnetic fields survive until the electroweak phase transition, they source a contribution to the baryon asymmetry of the universe. In this letter, we demonstrate that lepton flavour asymmetries above $|\mu|/T \sim 
9 \times 10^{-3}$ trigger this mechanism even for vanishing total lepton number. This excludes the possibility of such large lepton flavour asymmetries present at temperatures above $10^6$~GeV, setting a constraint which is about two orders of magnitude stronger than the current CMB and BBN limits.
\end{abstract}


\maketitle


\noindent\textbf{Introduction\,---\,}%
The observed baryon-to-photon ratio  $\eta_B^{\rm obs} = n_{\rm b}/n_\gamma = \left(6.12 \pm 0.04\right) \times 10^{-10}$~\cite{Planck:2018vyg,ParticleDataGroup:2020ssz}, together with the baryon-plus-lepton number ($B+L$) violating sphaleron processes in the Standard Model (SM), constrains the baryon and lepton number asymmetries in the thermal plasma of the early universe at temperatures above the electroweak phase transition (EWPT) to $|\mu_{B-L}|/T \lesssim 10^{-9}$.  The \emph{lepton flavour asymmetries (LFAs)}, carrying charge $\Delta_\alpha \equiv B/3 - L_\alpha$ 
with $\alpha = e, \mu, \tau$, could however be much larger as long as an (approximate) $B-L$ symmetry insures $|\sum_\alpha \mu_{\Delta_\alpha}/T| \lesssim 10^{-9}$. 
Taking into account neutrino oscillations which become efficient just before the onset of Big Bang Nucleosynthesis (BBN), the constraint on the asymmetry in the electron neutrinos at the time of BBN, 
$\mu_{\Delta_e} / T_\nu |_\text{BBN} = - 0.001 \pm 0.016$~\cite{Pitrou:2018cgg}, merely limits such primordial LFAs to $  |\mu_{\Delta_\alpha}|/T_\nu \lesssim 0.12 \, (1.0) \, g_{*,s}(T)/g_{*,s}^\text{BBN}$ 
for the two values for the neutrino mixing angle $\sin^2\theta_{13} = 0$  and $0.04$ considered in Refs.~\cite{Pastor:2008ti,Mangano:2010ei,Castorina:2012md}. 
Here $g_{*,s}$ accounts for the number of relativistic degrees of freedom at different temperatures.
The resulting contribution to extra radiation is at most around $\Delta N_\text{eff} \simeq 0.05$.
These bounds are considerably weaker than in the case of significant $B-L$ violation, $\mu_{B-L} \sim \mu_{\Delta_\alpha}$, for which the bounds on the electron-flavour asymmetry at BBN apply to all primordial LFAs
~\cite{Mangano:2010ei,Barenboim:2016lxv,Burns:2022hkq,Escudero:2022okz} (see~\cite{Iocco:2008va,Pitrou:2018cgg} for a review and e.g. \cite{Oldengott:2017tzj,Burns:2022hkq,Escudero:2022okz,Kumar:2022vee} for CMB constraints).

The possibility of such large LFAs has recently received renewed  attention, in particular as a possibility to explain the baryon asymmetry of our universe through leptoflavourgenesis~\cite{Mukaida:2021sgv} (see also Refs.~\cite{Kuzmin:1987wn,Khlebnikov:1988sr,March-Russell:1999hpw,Laine:1999wv,Shu:2006mm,Gu:2010dg} for related works) and as a possible explanation for the recently observed helium anomaly~\cite{Matsumoto:2022tlr,Burns:2022hkq}, indicating a smaller value for primordial helium-4 abundance compared to the standard BBN prediction (see e.g.~\cite{Kohri:1996ke,March-Russell:1999hpw,Pastor:2008ti} for earlier works). 
Lepton (flavour) asymmetries have moreover been considered to ameliorate the Hubble tension~\cite{Seto:2021tad} and improve the overall fit to cosmological data~\cite{Yeung:2020zde}.
See e.g.~\cite{Dreiner:1992vm,Casas:1997gx,McDonald:1999in,March-Russell:1999hpw,Kawasaki:2002hq,Yamaguchi:2002vw,Takahashi:2003db,Asaka:2005pn,Shu:2006mm,Gu:2010dg,Harigaya:2019uhf,Gelmini:2020ekg,Mukaida:2021sgv,Kawasaki:2022hvx} for models generating large lepton (flavour) asymmetries and their implications for baryogenesis.

In this letter we derive a new constraint on LFAs present in the early universe above a temperature of $10^6$~GeV, which is significantly stronger than existing constraints except for the special case of an (approximate) ${\mu + \tau}$ symmetry.
This new constraint will in particular rule out tauphobic leptoflavourgenesis from $\mu$ asymmetry and will equally rule out primordial LFAs (generated at $T>10^6$~GeV) 
 as a possible explanation to the helium anomaly.
The essence of this new constraint is the observation that LFAs can trigger a chiral plasma instability (CPI) which sources helical hypermagnetic fields~\cite{Joyce:1997uy} (see also \cite{Brandenburg:2017rcb,Schober:2017cdw,Schober:2018ojn}). These helical magnetic fields survive until the EWPT, at which their conversion into electromagnetic fields sources a contribution to the baryon asymmetry of the universe~\cite{Giovannini:1997gp,Giovannini:1997eg,Kamada:2016cnb}. Avoiding overproduction of the baryon asymmetry places an upper bound on the LFAs. 
Thus, in a similar spirit that non-perturbative $SU(2)_L$ processes (sphalerons) together with the observed baryon asymmetry set a constraint on $L$ and $B-L$ asymmetries, we point out that non-perturbative $U(1)_Y$ processes (CPI) constrain lepton flavour asymmetries.


\smallskip\noindent\textbf{Chiral plasma instability\,---\,}%
Hypermagnetic fields in the thermal plasma of the early universe can be described by chiral magnetohydrodynamics (MHD)~\cite{Durrer:2013pga},
\begin{align}
 0 = \frac{\partial \bm{B}_Y}{\partial \eta} + \bm{\nabla} \times \bm{E}_Y \,, \quad 0 = \bm{\nabla} \times \bm{B}_Y - \bm{J}_Y \,,
 \label{eq:MHD}
\end{align}
where $\eta$ denotes conformal time and
\begin{align}
 \bm{J}_Y = \sigma_Y (\bm{E}_Y + \bm{v} \times \bm{B}_Y) + \frac{2 \alpha_Y}{\pi} \mu_{Y,5} \bm{B}_Y \,.
 \label{eq:current}
\end{align}
Here $\sigma_Y \simeq 10^2 T$ denotes the conductivity of the thermal plasma, $\bm{v}$ is the fluid velocity, $\alpha_Y$ is the hypercharge fine structure constant of the hypercharge gauge group $U(1)_Y$ and $\mu_{Y,5}$ is the chiral chemical potential associated with $U(1)_Y$,
\begin{align}
 \mu_{Y,5} = \sum_{i} \varepsilon_i g_i Y_i^2 \mu_i \,,
 \label{eq:mu5def}
\end{align}
where $\varepsilon_i = \pm 1$ denotes right/left-handed particles, $g_i$ is the multiplicity and $Y_i$ is the hypercharge of the SM particle species $i$. The second term in Eq.~\eqref{eq:current}, referred to as the chiral magnetic effect~\cite{Vilenkin:1980fu,Alekseev:1998ds,Son:2004tq,Fukushima:2008xe}, is the origin of the chiral plasma instability~\cite{Joyce:1997uy}. It will prove convenient to express Eq.~\eqref{eq:MHD} in terms of the helicity stored in the hypermagnetic fields and the chiral chemical potential~\cite{Boyarsky:2011uy,Domcke:2019mnd},
\begin{align}
 \partial_\eta h_k & = - \frac{2 k^2}{\sigma_Y} h_k + \frac{ 
 4 \alpha_Y}{\pi} \frac{\mu_{Y,5}}{\sigma_Y} \rho_{B,k} \\
 \partial_\eta \rho_{B,k} & =  - \frac{2 k^2}{\sigma_Y} \rho_{B,k} + \frac{ 
 \alpha_Y}{\pi} \frac{\mu_{Y,5}}{\sigma_Y} k^2 h_{k} \,,
\end{align}
where $h_k$ and $\rho_{B,k}$ are the Fourier components of the hypermagnetic helicity and energy density, respectively, and the fluid velocity has been neglected. Combining these two equations, all modes $k < k_\text{CPI} \equiv 
\alpha_Y |\mu_{Y,5}|/\pi$ become tachyonically unstable, leading to the generation of helical hypermagnetic fields with a typical length scale of $1/k_\text{CPI}$ seeded by thermal fluctuations. The fastest growing mode is $k \sim k_\text{CPI}/2$ and the time scale of its growth can be estimated as $\eta_\text{CPI} \sim 2 \sigma_Y/k_\text{CPI}^2$, indicating that the CPI becomes effective at~\cite{Kamada:2018tcs} 
\begin{align}
  T_\text{CPI} \sim 10^5~\text{GeV} \,  \left(\frac{10^2}{g_*}\right)^{\tfrac{1}{2}} \left( \frac{\alpha_Y}{0.01} \right)^2 \left( \frac{10^2 T}{\sigma_Y}\right) \left( \frac{\mu_{Y,5}/T}{2 \cdot 10^{-3}} \right)^2\bigg|_{T_\text{CPI}} .
 \label{eq:TCPI}
\end{align}
This analytical estimate is in good agreement with the numerical MHD simulations presented in~\cite{Schober:2017cdw}.

We expect that thermal fluctuations provide initial seeds of hypermagnetic helicity of order $h_k \sim T^4 (k/T)^3/ k$ for $k \ll T$, where $(k/T)^3$ represents the suppression at the tail of the Bose-Einstein distribution. This should be amplified to $\mathcal{O}(T^2 |\mu_{Y,5}^{\rm ini}| / \alpha_Y)$ to complete the CPI, as we will see shortly, 
where $\mu_{Y,5}^\text{ini}$ denotes the value of the chiral chemical potential at the onset of the CPI.
Focusing on the fastest growing mode, we estimate the time scale of the completion of the CPI to be $\eta_{\rm CPI} * \ln \alpha^{-4} (T/\mu_{Y,5}^{\rm ini})^2 \sim \mathcal{O}(10) \eta_{\rm CPI}$. 
The chiral plasma instability ends once $\mu_{Y,5} \simeq 0$, i.e.\ when the chiral asymmetry in the plasma has been converted to helical magnetic fields.%
\footnote{
In practice, it suffices that $|\mu_{Y,5}| \lesssim 10^{-3}$ to end the CPI, since this pushes $T_\text{CPI}$ below the equilibration temperature of the electron Yukawa, which will efficiently complete the erasure of $\mu_{Y,5}$ as discussed below.
}
At the final stages of the CPI the effect of the velocity fields can no longer be neglected, 
but the main conclusions drawn above remain valid~\cite{Schober:2017cdw}.


\smallskip\noindent\textbf{Conserved charges in the SM plasma\,---\,}%
Besides the four well-known conserved charges of the SM above the electroweak phase transition (hypercharge and the three flavoured $B-L$ charges $\Delta_\alpha$) the SM plasma in the early universe also features approximately conserved charges whenever Yukawa couplings or non-perturbative sphaleron processes are not efficient enough to keep up with the expansion rate of the universe. At any given temperatures, approximating the SM interactions to be either inefficient or equilibrated,
the chiral chemical potential~\eqref{eq:mu5def} can be expressed as a linear combination of the respective conserved charges, with all other SM chemical potentials entering Eq.~\eqref{eq:mu5def} expressed in terms of these conserved charges~\cite{Domcke:2020kcp}. 

Our main focus in this letter will be on the temperature regime $10^9~\text{GeV} \gtrsim T \gtrsim 10^6$~GeV, where the weak and strong sphaleron process as well as all Yukawa couplings of the second and third generation are efficient. The Yukawa couplings of the first generation quarks, as well as the electron Yukawa coupling and the off-diagonal down-strange quark Yukawa coupling remain inefficient, conserving the charges associated with $\mu_{u-d}$, $\mu_e$ and $\mu_{2B_1 - B_2 - B_3}$. Solving the system of linear equations for all chemical potentials including the equilibrated SM interactions Domcke:2022kfs equations (see~\cite{Domcke:2020kcp,Domcke:2022kfs} for details) yields
\begin{align}
 \frac{\mu_{Y,5}}{T} =   \frac{513}{358} \frac{\mu_e}{T}  + \frac{173}{1074} \frac{\bar \mu_{u-d}}{T}  + \frac{151}{358} \frac{\mu_{\Delta_e}}{T} - \frac{10}{179}   \frac{\mu_{\Delta_{\mu + \tau}} }{T} \,,
 \label{eq:mu5_1}
\end{align}
for  $10^9~\text{GeV} \gtrsim T \gtrsim 10^6$~GeV. 
Here the bar indicates that we have summed over the three color degrees of freedom of the $u-d$ charge and $\mu_{\Delta_{\mu + \tau}} \equiv \mu_{\Delta_\mu} + \mu_{\Delta_\tau}$.
In the remainder of this letter we will for simplicity assume initial conditions with $\mu_e^\text{ini} =  \bar \mu_{u-d}^\text{ini} = 0$ and $\sum_\alpha \mu_{\Delta_\alpha} = 0$. Eq.~\eqref{eq:mu5_1} demonstrates that a $B$$-$$L$ flavour asymmetry generically generates a non-vanishing value for the chiral chemical potential $\mu_{Y,5}$ at  $10^9~\text{GeV} \gtrsim T \gtrsim 10^6$~GeV. 

As described above, such a non-zero $\mu_{Y,5}$ can trigger a CPI which drives $\mu_{Y,5}$ to zero, at the cost of generating a fermion asymmetry as well as generating helical hypermagnetic fields. The equations for the individual fermion currents $J_i^\mu$ are dictated by the chiral anomaly,
\begin{align}
 \partial_\mu J^\mu_i = \varepsilon_i g_i Y_i^2 \frac{\alpha_Y}{ \pi} \bm{E}_Y \bm{B}_Y + \dots \,,
\end{align}
where the dots indicate the SM Yukawa interaction and sphaleron processes and the zero component of the current is determined by the corresponding chemical potential, $q_i = \bar \mu_i T^2/6$. Given that in the temperature range of interest, these do not affect the $e$ and $u-d$ currents, the charge associated with the linear combination $\mu_e - \bar \mu_{u-d} = 0$ is preserved throughout the CPI. Together with setting $\mu_{Y,5} = 0$ at the completion of the CPI in Eq.~\eqref{eq:mu5_1}, we obtain
\begin{align}
  \frac{856}{537} \frac{\mu_e}{T}  =  - \frac{151}{358} \frac{\mu_{\Delta_e}}{T} + \frac{10}{179}  \frac{\mu_{\Delta_{\mu+\tau}}}{T}  
   = - \frac{\mu_{Y,5}^\text{ini}}{T}\,,
  \label{eq:mu5_2}
\end{align}
right after the CPI has completed. 
The conservation law for the total helicity density, derived from the chiral anomaly equation, then dictates the generation of helicity density
\begin{align}
 h  = - \frac{\pi T^2}{3 \alpha_Y} \mu_e = - \frac{\pi T^2}{3 \alpha_Y} \bar\mu_{u-d}
   = \frac{\pi T^2}{\alpha_Y} \, 
   \frac{179}{856} \, \mu_{Y,5}^\text{ini} \,, 
    \label{eq:helicity}
\end{align}
where $\mu_e$ and $\bar \mu_{u-d} = \mu_e$ denote the asymmetry in the right-handed electrons and first generation quarks after the CPI and we have assumed zero initial net helicity.

When the temperature drops below $10^6$~GeV, the first generation quark Yukawa couplings equilibrate and $\bar \mu_{u-d}$ is no longer associated with a conserved charge. Eq.~\eqref{eq:mu5_1} is replaced by
\begin{align}
 \frac{\mu_{Y,5}}{T} =   \frac{711}{481} \frac{\mu_e}{T}   + \frac{5}{13} \frac{\mu_{\Delta_e}}{T} - \frac{4}{37}   \frac{\mu_{\Delta_{\mu + \tau}} }{T} \,,
 \label{eq:mu5_3}
\end{align}
which, when compared to Eq.~\eqref{eq:mu5_2} and taking into account $\mu_{\Delta_{\mu+\tau}} = - \mu_{\Delta_e}$, only marginally modifies the final value for $\mu_e$ and hence the helicity if the CPI occurs in this temperature range.%
\footnote{For completeness, we note that in the temperature regime $10^{11}~\text{GeV} > T > 10^9$~GeV, when the muon Yukawa and some of the second and third generation quark Yukawa couplings are not equilibrated, the analogue of Eq.~\eqref{eq:mu5_2} reads 
\begin{align}
 \frac{\mu_{Y,5}}{T} =   \frac{1765}{589} \frac{\mu_e}{T}   + \frac{188}{589} \frac{\mu_{\Delta_{e+\mu}}}{T} - \frac{88}{589}   \frac{\mu_{\Delta_{\tau}} }{T} \,, \nonumber
\end{align}
with coefficients which are numerically again quite similar to Eq.~\eqref{eq:mu5_2}. Note however that since only the third generation lepton Yukawa coupling is in equilibrium, $\mu+\tau$ symmetric LFAs yield a non-vanishing $\mu_{Y,5}$ whereas the $e + \mu$ symmetric case does not.}
At $10^5$~GeV the electron Yukawa interaction equilibrates~\cite{Bodeker:2019ajh}, $\mu_e$ becomes a function of $\mu_{\Delta_\alpha}$, and $\mu_{Y,5}$ vanishes independent of the initial values for $\mu_{\Delta_\alpha}$. Hence the CPI can only be triggered above the electron Yukawa equilibration temperature of about $10^5$~GeV. Taking into account the discussion below Eq.~\eqref{eq:TCPI}, this means that the CPI should become effective at a temperature above $\mathcal{O}(10^6) \GeV$ in order to complete by the time of $T = \mathcal{O}(10^{5}) \GeV$.


\smallskip\noindent\textbf{Baryogenesis from decaying helical magnetic fields\,---\,}%
If this helicity survives until the EWPT, then the conversion of hypermagnetic field to electromagnetic field generates a baryon asymmetry~\cite{Kamada:2016cnb},
\begin{align}
 \eta_B^0 =  c_B^\text{dec} \frac{\alpha_Y}{2 \pi} \frac{h}{n_\gamma} \left(\frac{g_{*,s}^0}{g_{*,s}^\mathrm{ewpt}} \right) \,.
 \label{eq:would-be_AS}
\end{align}
Here $g^0_{*,s}/g_{*,s}^\mathrm{ewpt} \simeq 0.04$, denotes the ratio of the degrees of freedom in the thermal plasma at the EWPT and today, $n_\gamma = 2 \,  \zeta(3) \, T^3/\pi^2$ is the photon number density and $c_B^\text{dec} \simeq 0.05$ parametrizes the efficiency of baryogenesis from decaying hypermagnetic fields at the EWPT~\cite{Kamada:2020bmb,Domcke:2022kfs}. Given current uncertainties on the dynamics of the EWPT, $c_B^\text{dec}$ may vary by almost three orders of magnitude~\cite{Kamada:2016cnb,Jimenez:2017cdr}. This does however not change the conclusion that any value $|\mu_{Y,5}^\text{ini}|/T \gtrsim 10^{-3}$ which is sufficient to trigger (and complete) the CPI before the equilibration of the electron Yukawa interaction, see Eq.~\eqref{eq:TCPI}, will lead to an baryon asymmetry which is orders of magnitude above the observed value of $\eta_B^\text{obs} \sim 10^{-9}$. This can be seen immediately by inserting Eq.~\eqref{eq:helicity} into Eq.~\eqref{eq:would-be_AS}.\footnote{
LFAs can also directly generate a baryon asymmetry during sphaleron decoupling, see e.g.~\cite{Kuzmin:1987wn,Khlebnikov:1988sr,March-Russell:1999hpw,Laine:1999wv,Mukaida:2021sgv}. This contribution is expected to be significantly smaller than the one obtained from Eq.~\eqref{eq:would-be_AS} and does not change our conclusions.
}
Moreover, 
our conclusions hold even if the electroweak phase transition is first-order due to beyond-the-SM effects, in which case the efficiency factor $c_B^{\rm dec}$ would be {\it larger}~\cite{Giovannini:1997eg}.

Such large values of the chiral chemical potential, and consequently large values of the helicity density, also ensure that the turbulent regime of MHD is reached, triggering a so-called cascade pushing the helicity to larger length scales and thus protecting it from magnetic diffusion operating at small scales~\cite{Durrer:2013pga,Kahniashvili:2012uj,Banerjee:2004df}. An estimate of the kinetic and magnetic Reynolds numbers returns values much larger than unity, indicating that a helicity generated at $10^9~\text{GeV} \gtrsim T \gtrsim 10^5$~GeV should indeed survive until the EWPT.


\smallskip\noindent\textbf{Constraints on LFAs\,---\,}%
From the discussion above we conclude that lepton flavour asymmetries $\mu_{\Delta_\alpha}$ which are large enough to generate a chiral chemical potential $\mu_{Y,5}$ which triggers and completes the CPI before the equilibration of the electron Yukawa coupling are excluded since they would overproduce the matter--antimatter asymmetry of the universe. Accounting for uncertainties in the determination of the onset of the CPI we consider the parameter space in which our estimate~\eqref{eq:TCPI} of the CPI temperature lies above $10^6$~GeV to be excluded. Combining Eq.~\eqref{eq:TCPI} and \eqref{eq:mu5_2} then yields
\begin{align}
 \left| \frac{151}{358} \frac{\mu_{\Delta_e}}{T} - \frac{10}{179} \left(  \frac{\mu_{\Delta_\mu}+\mu_{\Delta_\tau}}{T} \right)  \right| < 
 4.1 \cdot 10^{-3} \,,
 \label{eq:bound_0}
\end{align}
where we have set $g_* = 106.75$, $\alpha_Y = 0.011$ and $\sigma_Y = 
50~T$ at $10^6$~GeV~\cite{Baym:1997gq,Arnold:2000dr}.
Imposing $B-L$ conservation, this translates to
\begin{align}
 \left| \frac{\mu_{\Delta_e}}{T} \right| =  \left| \frac{\mu_{\Delta_\mu} + \mu_{\Delta_\tau}}{T} \right| < 
 8.7
 \cdot 10^{-3} \,,
 \label{eq:bound}
\end{align}
which is the main observation of this letter. 

To put this constraint into context, let us summarize the assumptions in our analysis and their impact on this result. This constraint applies to LFAs present before the onset of the CPI, notably at temperatures of the SM thermal bath above $10^5$~GeV.
To obtain~\eqref{eq:bound_0} we have moreover taken all other asymmetries, in particular the asymmetry in the right-handed electrons and a possible helical hypermagnetic background field to be zero at the onset of the CPI. A violation of the latter condition will change the numerical factors in~\eqref{eq:bound_0} but will generically yield a comparable bound. A notable exception to this is if the net asymmetry stored in the fermion chemical potentials and in the helical gauge fields vanishes, as is the case in axion inflation~\cite{Domcke:2018eki}. In this case, the CPI erases all asymmetries in the system and the constraint~\eqref{eq:bound_0} disappears.

To compare our result with the existing bounds in the literature, we have to account for the entropy injection by the decoupling of relativistic particles. Noting that $ T^2 \mu_{\Delta_\alpha} / s$ is preserved in an adiabatically expanding universe,  with $s$ denoting the entropy density,
we obtain
\begin{align}
 \left. \frac{\mu_{\Delta_\alpha}}{T} \right\vert_{T = T_1} = \left( \frac{g_{*,s}(T_1)}{g_{*,s}(T_2)} \right) \left. \frac{\mu_{\Delta_\alpha}}{T} \right\vert_{T = T_2} \,,
\end{align}
where in particular $g_{*,s}^{\rm BBN}/g_{*,s}^{\rm ewpt} \simeq 0.1$. 
This in particular provides a bound on the LFAs which is about two orders of magnitude stronger than existing bounds on primordial lepton flavour asymmetries~\cite{Pastor:2008ti,Mangano:2010ei}. In fact, inserting $\mu_{\Delta_e}/T_\nu = - (\mu_{\Delta_\mu} + \mu_{\Delta_\tau})/T_\nu$ with $ |\mu_{\Delta_e}|/T_\nu \lesssim 1$ 
into Eq.~\eqref{eq:mu5_1} yields $\mu_{Y,5} \lesssim 0.5$ and thus $T_\text{CPI} \lesssim 10^{10}$~GeV, justifying our focus on the temperature range of $10^9~\text{GeV} \gtrsim T \gtrsim 10^6$~GeV for the onset of the CPI. 
Moreover, our constraint excludes tauphobic leptoflavourgenesis, which considers $\mu_{\Delta_\mu}/T = - \mu_{\Delta_e}/T \simeq 0.4$ and $\mu_{\Delta_\tau}/T = 0$~\cite{March-Russell:1999hpw,Mukaida:2021sgv}, 
if the  
asymmetries are generated above $10^6$~GeV. 
On the other hand, leptoflavourgenesis with a sizable tau flavour component, 
$\mu_{\Delta_\tau}/T \simeq 8 \cdot 10^{-3}$~\cite{Shu:2006mm,Gu:2010dg,Mukaida:2021sgv} is marginally consistent with our bound within the uncertainty that comes from the rough estimation for the time scale of completion of the CPI ($=\mathcal{O}(10) \eta_{\rm CPI}$). 

A large asymmetry in the electron flavour,  $- \mu_{\Delta_e}/T_\nu = \mu_{\nu_e}/T_\nu \simeq 0.04$, has been proposed e.g.\ in~\cite{Burns:2022hkq} to address the helium anomaly. One possible implementation of this is a significant violation of $B$$-$$L$ after the EWPT but before BBN, resulting in $\mu_{\nu_\mu}/T_\nu \simeq \mu_{\nu_\tau}/T_\nu \sim \mu_{\nu_e}/T_\nu \simeq 0.04$ at BBN, see e.g.~\cite{Borah:2022uos}. Alternatively, if the LFAs are created before the EWPT, $|\mu_\text{B-L}|/T \lesssim 10^{-9}$ together with the equilibration of LFAs through neutrino oscillations just before the onset of BBN, leads to a significant suppression of the impact of LFAs on BBN and CMB observations~\cite{Froustey:2021azz}. This is particularly relevant given the relatively large neutrino mixing angle $\sin^2 \theta_{13} = 0.022$~\cite{ParticleDataGroup:2020ssz} which leads to an onset of the electron neutrino oscillations before BBN.  As demonstrated in~\cite{Pastor:2008ti,Mangano:2010ei}, the neutrino distributions do however not reach full kinetic equilibrium before decoupling, and the resulting deviation from a Fermi-Dirac distribution leads to non-vanishing effective values of $\mu_{\nu_\alpha}^\text{eff}/T_\nu$ which impact both the light element abundances produced during BBN as well as the surviving neutrino radiation $\Delta N_\text{eff}$. Obtaining $\mu_{\nu_e}^\text{eff}/T_\nu \simeq 0.04$ to address the helium anomaly, requires a primordial value of $-\mu_{\Delta_e}/T_\nu = (\mu_{\Delta_\mu} + \mu_{\Delta_\tau})/T_\nu = {\cal O}(1)$ at $T \sim 10 \MeV$~\cite{Pastor:2008ti,Mangano:2010ei}, which is firmly ruled out by our new constraint~\eqref{eq:bound}. Our constraint moreover excludes the possibility that the helium anomaly is addressed by a more moderate LFA, $-\mu_{\Delta_e}/T_\nu \simeq 0.04$, with the onset of neutrino oscillations delayed by non-standard neutrino interactions~\cite{Dolgov:2004jw}. We conclude that our  new constraint~\eqref{eq:bound} rules out the possibility of explaining the helium anomaly with primordial LFAs, independent of the precise equilibration temperature of the neutrino oscillations.

Two obvious caveats to this constraint deserve to be mentioned. First, if the LFAs are generated only at temperatures below $10^5$~GeV, the constraints derived here do not apply. 
Scenarios considered in Refs.~\cite{Dreiner:1992vm,Casas:1997gx,McDonald:1999in,Kawasaki:2002hq,Yamaguchi:2002vw,Takahashi:2003db,Asaka:2005pn,Harigaya:2019uhf,Gelmini:2020ekg,Kawasaki:2022hvx} are in this category because they generate large lepton (flavour) asymmetry after the electroweak phase transition. Second, in models with $\mu + \tau$ symmetry (in addition to the total $B$$-$$L$ symmetry), the chiral chemical potential $\mu_{Y,5}$ vanishes below $10^9$~GeV and the constraints derived here are evaded. Note that in this latter case the LFAs are erased once $\mu - \tau$ neutrino oscillations begin, making a solution to the helium anomaly based on this construction challenging.


\smallskip\noindent\textbf{Conclusions\,---\,}%
In this letter we point out that non-perturbative SM processes associated with the chiral magnetic effect in the hypercharge gauge group can be used to set constraints on large lepton flavour asymmetries present in the early universe at temperatures above a $10^6$~GeV. In the absence of a $\mu + \tau$ symmetry, we constrain the flavoured $B$$-$$L$ asymmetries to $|\mu_{\Delta_\alpha}|/T <  0.009$. These constraints are currently not limited by experimental accuracy, but rather by theory uncertainties. A more accurate simulation of the dynamics of the chiral plasma instability in the regime close to the equilibration temperature of the electron Yukawa interaction could potentially improve this bound by a factor $\sqrt{10}$ by resolving the regime where the CPI becomes relevant but is not completed before the electron Yukawa equilibrates. 
In this regime, it may moreover be possible to obtain the observed baryon asymmetry, as discussed in Ref.~\cite{Kamada:2018tcs}.
Further progress may be made by dropping the approximation of instant equilibration of the various Yukawa couplings and instead solving the Boltzmann equations for the Yukawa interactions once they become marginally relevant. We hope that our work sparks future research in these directions.

While the focus of this letter is on constraining lepton flavour asymmetries, the mechanism considered here also constrains scenarios where any of the fermion asymmetries is large, even if the asymmetry is washed out at lower temperatures (see e.g.~\cite{Co:2019wyp,Co:2020xlh,Co:2020jtv,Co:2019jts}). This also includes, e.g., scenarios of leptoflavourgenesis that rely on large fermionic input charges generated at very high energies. The transport equations of the SM will redistribute the asymmetries according to the conserved charges in the different temperature regimes, but generically at temperatures above $10^5$~GeV, $\mu_{Y,5}$ is of the same order as the largest initial fermion asymmetry  (see, e.g., Ref.~\cite{Domcke:2020kcp}). As discussed in this letter, this can trigger the CPI, generating helical magnetic fields which can lead to an overproduction of the baryon asymmetry.


\medskip\noindent\textit{Acknowledgments\,---\,}%
We thank Miguel Escudero for helpful discussions on the helium anomaly as well as Keisuke Harigaya and Mikhail Shaposhnikov for comments on the draft.
K.~K. was supported by JSPS KAKENHI, Grant-in-Aid for Scientific Research (C) JP19K03842. 
K.\,M.\, was supported by MEXT Leading Initiative for Excellent Young Researchers Grant No.\ JPMXS0320200430,
and by JSPS KAKENHI Grant No.\ 	JP22K14044.
M.\,Y.\ was supported by the Leading Initiative for Excellent Young Researchers, 
MEXT, Japan, and by JSPS KAKENHI Grant No.\ JP20H05851 and JP21K13910.
%

\bibliographystyle{JHEP}
\bibliography{refs}


\newpage
\onecolumngrid
\newpage


\renewcommand{\thesection}{S\arabic{section}}
\renewcommand{\theequation}{S\arabic{equation}}
\renewcommand{\thefigure}{S\arabic{figure}}
\renewcommand{\thetable}{S\arabic{table}}
\setcounter{equation}{0}
\setcounter{figure}{0}
\setcounter{table}{0}
\setcounter{page}{1}





\end{document}